# Electrically-Insulating Flexible Films with Quasi-One-Dimensional van-der-Waals Fillers as Efficient Electromagnetic Shields

Zahra Barani[1], Fariborz Kargar[1], Yassamin Ghafouri[2], Subhajit Ghosh[1], Konrad Godziszewski[3], Saba Seyedmahmoudbaraghani[1,6], Yevhen Yashchyshyn[3,4], Grzegorz Cywiński[4,5] Sergey Rumyantsev[4], Tina T. Salguero[2] and Alexander A. Balandin[1,6*]

[1]Nano-Device Laboratory (NDL) and Phonon Optimized Engineered Materials (POEM) Center, Department of Electrical and Computer Engineering, University of California, Riverside, California 92521 USA

[2]Department of Chemistry, University of Georgia, Athens, Georgia 30602 USA

[3]Institute of Radioelectronics and Multimedia Technology, Warsaw University of Technology, Warsaw 00-665 Poland

[4]CENTERA Laboratories, Institute of High-Pressure Physics, Polish Academy of Sciences, Warsaw 01-142 Poland

[5]CEZAMAT, Warsaw University of Technology, 02-822 Warsaw, Poland

[6]Materials Science and Engineering Program, University of California, Riverside, California 92521 USA

* Corresponding author (A.A.B.): balandin@ece.ucr.edu ; web-site: http://balandingroup.ucr.edu/

## Abstract

We report polymer composite films containing fillers comprised of quasi-one-dimensional (1D) van der Waals materials, specifically transition metal trichalcogenides containing 1D structural motifs that enable their exfoliation into bundles of *atomic threads*. These nanostructures are characterized by extremely large aspect ratios of up to $\sim 10^6$. The polymer composites with low loadings of quasi-1D $TaSe_3$ fillers (<3 vol. %) revealed excellent electromagnetic interference shielding in the X-band GHz and EHF sub-THz frequency ranges, while remaining DC *electrically insulating*. The unique electromagnetic shielding characteristics of these films are attributed to effective coupling of the electromagnetic waves to the high-aspect-ratio electrically-conductive $TaSe_3$ atomic-thread bundles even when the filler concentration is below the electrical percolation threshold. These novel films are promising for high-frequency communication technologies, which require electromagnetic shielding films that are flexible, lightweight, corrosion resistant, electrically insulating and inexpensive.

The explosive growth of interest in two-dimensional (2D) layered van der Waals materials, such as graphene and transition metal dichalcogenides (TMDs) $MX_2$, where M = transition metals and X = S, Se, Te, has resulted in numerous breakthroughs in physics and is expected to lead to important practical applications.[1–5] Recently, a different group of layered van der Waals materials with quasi-one-dimensional (1D) crystal structures has attracted significant attention: the transition metal trichalcogenides (TMTs). These compounds contain 1D motifs comprised of $MX_3$ atomic chains that are weakly bound together by van der Waals forces or chalcogen interactions. Examples of such materials include $TiS_3$, $NbS_3$, $TaSe_3$, and $ZrTe_3$.[6–13] $MX_3$ materials exfoliate into nanowire- and nanoribbon-type structures, as opposed to atomic planes of quasi-2D van der Waals $MX_2$ materials. We previously discovered that bundles of quasi-1D $TaSe_3$ atomic threads and $ZrTe_3$ nanoribbons can support high current densities of $J_B \sim 10$ $MA/cm^2$ and $J_B \sim 100$ $MA/cm^2$, respectively.[6,7,14] In this group of $MX_3$ materials, $TaSe_3$ is particularly interesting. Reported studies generally agree that it is a metal, with superconducting properties at low temperature, although some studies suggest that it is semimetal.[15–18] The exact band and specifics of electron transport in bulk and exfoliated nanowires of $TaSe_3$ at various temperatures are still under intensive debate.

In this Letter, we demonstrate that quasi-1D van der Waals materials like $TaSe_3$ can be used as the high-aspect-ratio metallic fillers in polymer composites to provide important functionality—efficient electromagnetic interference (EMI) shielding—in a wide frequency range that is relevant to current 5G and future communication technologies: X-band ($f$=8.2 GHz – 12.4 GHz) and the extremely high frequency (EHF) band ($f$=220 GHz – 320 GHz). Proliferation of portable devices and wireless communications has led to problems with environmental electromagnetic pollution. There is a need for more efficient EMI shielding materials characterized by low-weight, mechanical stability, resistance to oxidation, flexibility, and ease of manufacturing. Many applications also require EMI shielding films to be electrically insulating to avoid short circuiting of electronic components. The conventional materials for EMI shielding are metals, which are utilized as coatings and enclosures.[19] Metals have charge carrier concentrations that enable them to block EM waves mostly by reflection. However, metallic EMI shields are heavy and prone to oxidation. An alternative approach to EMI shielding is based on the use of polymers containing electrically-conductive fillers.[20–23] The first generation of polymer composites for EMI shielding utilized large loading fractions of metallic particles, such that their concentrations are above the electrical *percolation threshold*, resulting in overall electrically conductive films. The high loading fraction of metallic fillers is required to provide sufficient EMI shieling at a given thickness of the film. Recently, attention has turned to carbon allotrope fillers, including carbon nanotubes and graphene.[24–29] For example, we demonstrated efficient EMI shielding in the wide GHz and sub-THz frequency ranges with high-loading graphene composites.[30,31] The advantages of quasi-2D graphene fillers include low weight, high thermal stability, anti-corrosive properties, and low cost at mass production. Here we demonstrate that quasi-1D van der Waals materials can be used as efficient fillers for EMI shielding that, in certain aspects, outperform their quasi-2D counterparts.

For experiments with quasi-1D fillers in electrically insulating films for EM shielding we selected TaSe$_3$. This quasi-1D van der Waals material is well-suited for this application due to its metallic electronic structure and good stability with respect to oxidation. The fact that this material revealed extremely high current densities when exfoliated into the bundles of atomic threads was an additional important factor.[6] Recent interest in TaSe$_3$ has included studies of its topological phases,[17,32–34] the effect of strain on its metallic *vs.* semiconducting states,[35] low temperature charge density wave states,[18] and our own work characterizing its current carrying capacity and low-frequency electronic noise.[36] The room temperature, monoclinic crystal structure of TaSe$_3$ (Fig. 1a) exhibits aligned chains of trigonal prismatic [TaSe$_6$] units oriented along the *b*-axis.[37] These chains are assembled into corrugated bilayers (sets of blue and purple chains in Fig. 1a) through Ta$^{\cdots}$Se interactions between adjacent chains. Neighboring bilayers are separated by van der Waals gaps.

The bulk, crystalline TaSe$_3$ used in this work was prepared by iodine-mediated chemical vapor transport (CVT) from the elements. A temperature gradient of 750 to 650 °C (source zone – growth zone) led to efficient crystal growth during a 2-week period. TaSe$_3$ crystals grew preferentially along the *b*-axis, leading to ribbon- or needle-like, filamentary morphologies ranging from less than one micron to ten of microns in width (Fig. 1b and 2c). The scanning electron microscopy (SEM) image in Fig. 1b shows a TaSe$_3$ crystal that was freshly exfoliated for energy dispersive spectroscopy (EDS) analysis. EDS mapping shows excellent overlap of Ta and Se (Fig. 1c), and quantitative EDS averaged across the entire mapped area (SI) provides a slightly Se-deficient composition of TaSe$_{2.85}$ similar to other samples prepared by CVT.[6,34] Powder x-ray diffraction (XRD) is consistent with the anticipated monoclinic TaSe$_3$ structure, and it confirms the phase purity of the as-prepared material (Fig. 1d).

[Figure 1: Characterization data for as grown material]

The preparation of composites with quasi-1D fillers involved chemical phase exfoliation and inclusion of high aspect-ratio exfoliated threads into three polymeric matrices of sodium alginate (SA), epoxy, and a special type of UV-light cured polymer (UV-polymer) as the base. The TaSe$_3$ crystals were subjected to solvent-assisted exfoliation separately in two different solvents of acetone and dimethylformamide (DMF). During this process, the bulk TaSe$_3$ was dispersed and exfoliated in the solvents using low power ultrasonic bath sonication. The dispersion was centrifuged to isolate the solids, and the procedure of sonication/centrifugation was repeated several times. A photograph of the resultant dispersion is shown in Figure 2d. The SEM image in Figure 2e shows the size and morphology of TaSe$_3$ nanowires post-exfoliation . The typical diameter of the exfoliated bundles of the atomic threads of TaSe$_3$ ranges from 50 nm to 100 nm while their length is in the range of several hundred micrometers. The fillers were mixed with UV-polymer, epoxy, and SA. The obtained flexible thin films and composites are shown in Figure 2f-h. Raman spectra of TaSe$_3$ were taken before and after exfoliation in different solvents, and after mixing with the polymer matrix to confirm the quality and stability of the quasi-1D TaSe$_3$ fillers. The measurements were performed in the backscattering configuration under $\lambda$ = 488 nm laser excitation using low power to prevent local heating. The spectrum displays characteristic peaks between 25 cm$^{-1}$ to 300 cm$^{-1}$, which originate from the primitive monoclinic structure of TaSe$_3$.[18,34,35] The peaks at 140 cm$^{-1}$, 164 cm$^{-1}$, 214 cm$^{-1}$ and 238 cm$^{-1}$ are assigned to the out-of-plane A$_{1g}$ phonon modes whereas the peaks at 176 cm$^{-1}$ and 185 cm$^{-1}$ to the B$_2$/A$_g$ modes. [35] The Raman data confirm the crystalline nature of the TaSe$_3$ filler and the preservation of its structural integrity after all processing steps.

[Figure 2: Characterization data of the EMI films with 1D fillers]

To determine EMI characteristics, we measured the scattering parameters, $S_{ij}$ (see Methods section). The scattering parameters define the EM coefficients of reflection, $R = |S_{11}|^2$, and transmission, $T = |S_{21}|^2$, which, in turn, allow one to calculate the coefficient of absorption, $A$, as $A = 1 - R - T$. A faction of the energy of EM wave, incident on the film, is reflected at the interface. The rest is absorbed inside the film or transmitted through it. Because part of EM energy is reflected from the interface, the coefficient of absorption, defined as the power percentile of the absorbed EM way in the medium to the total power of incident wave, is not truly indicative of material's ability in absorbing the EM waves. For this reason, the effective absorption coefficient, $A_{eff}$, is defined as $A_{eff} = (1 - R - T)/(1 - R)$. The total shielding efficiency, $SE_T$, describes the total attenuation of the incident EM wave by the material of interest. This parameter determines the material's ability to block the EM waves and consists of two terms – the reflection shielding efficiency, $SE_R$, and the absorption shielding efficiency, $SE_A$. These parameters can be calculated in terms of $R$, $T$, and $A_{eff}$ as follows $SE_R = -10\log(1 - R)$, $SE_A = -10\log(1 - A_{eff})$, and $SE_T = SE_R + SE_A$. The reflection, absorption, and the total EMI shielding efficiencies of the UV-cured flexible polymer films with low concentrations of TaSe₃ fillers are presented in Figure 3a-c. As one can see, a thin film with 130 µm thickness and an extremely low concentration of 1.14 vol% of quasi-1D TaSe₃ fillers reveal strong EMI shielding of ~10 dB, *i.e.* 90% of the incident EM power on the film is shielded via reflection at the air-composite interface or absorption as it passes through the composite. Typically, the EMI shielding efficiency increases with the increasing filler loading.

[Figure 3: EMI characteristics of the films with 1D fillers in X-band frequency range]

One can see from Figure 3 that 15 dB of total EMI shielding can be achieved in 77-µm thick SA-based films with only 2.8 vol. % of quasi-1D van der Waals fillers. The shielding due to absorption of EM waves is approximately twice as much as that due to reflection of the waves. The fact that the shielding is achieved mostly via absorption versus reflection is beneficial for many practical applications that require minimization of the EM energy redirected back to the environment. The total EMI shielding efficiency $SE = SE_T$, which indicates how much EM energy is blocked by a film of particular thickness, is not the only characteristic that has to be considered for practical purposes. Another commonly used metric is the efficiency normalized by the mass density, $SSE = SE_T/\rho$. However, $SSE$ does not fully describe the EMI shielding of a given material because, by increasing the thickness of a film at a constant mass density, one can achieve higher and higher $SSE$ values. To better describe the EMI shielding at the material level, one can normalize SSE by the thickness, $t$, and use $SSE/t$ to compare the effectiveness of different composites.[21] Here we argue that, for many practical purposes, it is meaningful to normalize $SSE/t = SE/(\rho \times t)$ by the loading fraction of the fillers. Achieving higher EMI shielding in the composite with the lowest loading of the fillers makes sense from the weight and cost considerations as well as for maintaining electrical insolation of the composite when required. Indeed, if one composite can deliver the required SE with a low loading of lightweight fillers while another needs 90 % loading of silver (Ag) particles, it is clear that the Ag composite likely will be heavy, expensive and electrically conductive.

To assess the performance of the polymeric composites with fillers, we define the figure-of-merit $Z_B = SE/(\rho \times t \times m_f)$ by introducing normalization by the mass fraction of the fillers $m_f = M_F/(M_B + M_F)$, where $M_F$ and $M_B$ are the masses of the filler and base polymer, respectively. It is interesting to note that the physical meaning of the $Z_B$ figure-of-merit is the total shielding efficiency of the films per the areal density of the fillers, *i.e.* $Z_B = SE/(M_F/A)$, here $A = V/t$ is the area of the sample of the volume $V$ and thickness $t$ (see Supplemental for the details of the derivation). The defined metric put more emphasis on the material performance, and specifically the filler performance. Figures 4a and 4b show the $SSE/t$ and $Z_B$ for several polymer composites with different fillers. One can see that our composites with quasi-1D van der Waals fillers outperforms composites with carbon nanotubes and graphene. Although a composite with Ag has better performance in terms of $SSE/t$, the composites with quasi-1D fillers exhibit superior $Z_B$ efficiency. The latter means that the polymer composites low areal density of quasi-1D fillers are extremely effective in blocking EM waves.

[Figure 4: EMI shielding characteristics of polymeric composites with different fillers]

As the next step, we examined the EMI shielding effectiveness of the composite with the low loading of quasi-1D TaSe$_3$ fillers in the EHF band. The measurements were performed using the quasi-optical free space method. The details of the experimental procedures are described in the Methods section. The coefficients of reflection, absorption, effective absorption, and transmission presented in Figures 5 (a) – (d) demonstrate the superior performance of the flexible films with quasi-1D fillers in the EHF band. Note that only 0.0002% of the incident EM wave is transmitted through the 1-mm-thick film with only 1.3 vol. % of quasi-1D TaSe$_3$ fillers. Figures 5 (b) and (c) indicate clearly that absorption is the dominant mechanism of the EMI shielding in EHF range. This is different from the situation in the X-band where the reflection was substantial. The absorption $SE_A$ increases from 55 dB to 76 dB as the frequency varies from 220 GHz to 320 GHz. The EMI shielding by reflection contributes only ~1.5 dB to the total shielding, and it slightly decreases from 1.7 dB to 1.4 dB as the frequency increases. For comparison, the inset shows pristine epoxy's characteristics in the same frequency range. As expected, epoxy by itself is a poor shielding material and provides only the mean value of $SE_T$~1.5 dB in the EHF range. The EMI shielding performance of quasi-1D fillers in insulating epoxy films is exceptional as compared to other fillers.

[Figure 5: EMI characteristics of the epoxy composite with 1D fillers in EHF band]

An important feature of the synthesized films is their electrical insulation. We verified that the DC electrical conductivity of the films is below the instrumentation measurement limit. The upper bound of the electrical resistivity is $10^{15}$ $\Omega$-cm. This means that the loading fraction of the quasi-1D van der Waals fillers is below the percolation threshold. This is rather surprising because, according to the conventional theories developed for carbon nanotubes and other materials with high aspect ratio, the electrical percolation should be attained at even lower loading <1 vol. %. [38–42] The disagreement with the known models can be explained by the fact that the conventional theories used the mathematical approximation of the fillers as straight cylinders, whereas we often observed TaSe$_3$ bending (see Figure 2), which could affect the percolation threshold. During material processing, we paid special attention to the uniformity of dispersion of the fillers, and verified it with microscopy. Some deviation from the uniformity over the sample total area is a possible factor, which requires a separate investigation.

Another interesting question is why electrically insulating films are so effective in blocking the EM waves. Even though the quasi-1D fillers do not create a percolated, electrically-conductive network, they effectively couple EM waves. The electric field of EM waves interacts with the free carriers in the quasi-1D conductors, and thus enables reflection and absorption of EM energy. One should also note that the frequency of 10 GHz, the EM wavelength $\lambda = c/(\varepsilon_r^{1/2} f) \sim 19$ mm (here $\varepsilon_r \sim 2.4$ is the relative dielectric constant of polymer base material). A few of connecting quasi-1D fillers with high aspect ratio would make something similar to an antenna, effective at receiving and re-emitting EM energy. A few connecting and bend quasi-1D fillers that form a circular loop would act similar to a magnetic antenna in this frequency range. These considerations can explain an efficiency of quasi-1D bundles of atomic threads as fillers in EMI shielding films. The "antenna" action in the X-band is consistent with the fact that reflection of EM waves made substantial contribution of the overall EMI shielding at these frequencies. In the EHF range, where the EM wavelength at $f$=300 GHz is $\lambda \sim 0.65$ mm, the randomly distributed quasi-1D fillers can act more as the scattering objects, which explain the dominance of absorption in the overall EMI shielding.

One should note that the electrical conduction properties of TaSe$_3$ itself are still not completely understood. Bulk TaSe$_3$ has not been studied in as much detail as other TMT materials, possibly due to its low superconducting phase transition T$_c$ ~2 K[15,43]. A variety of measurements indicate that TaSe$_3$ is metallic or semi-metallic down to T$_c$,[44–47]. At the same time, some reports suggested that stress or strain can produce a semiconducting gap.[35,48] Despite the fact that many published studies do not include compositional data (e.g., EDS, EPMA, ICP-MS/OES), at least some CVT-grown $TaSe_3$ crystals appear selenium deficient,[18,49] approximately $TaSe_{2.8}$, like the ones used in this work. Surprisingly, selenium-deficient $TaSe_3$ can be produced from even selenium-rich CVT conditions.[34,50,51] Stoichiometric $TaSe_3$ has been isolated from high pressure conditions and selenium-flux growth.[33,52] Although selenium deficiency does not seem to affect the overall electrical conductivity of $TaSe_3$ or its $T_c$,[15,37,43,45–47] several studies indicate that doping can modify its electronic structure. For example, the mixed chalcogenide $Ta(S_xSe_{1-x})_3$ becomes semiconducting with increasing sulfur content,[53] and indium impurities from contacts to $TaSe_3$ can produce a metal-insulator transition.[54,55] In addition, the intercalation of copper into $TaSe_3$ causes $T_c$ lowering and weak induced CDWs.[56] Further investigations clearly are needed to understand the impact of defects and dopants on the electrical properties of $TaSe_3$.

In conclusion, we demonstrated that quasi-1D van der Waals materials can be used as fillers in flexible polymer films providing excellent EMI shielding capability in the X-band and EHF-band. Polymer composites films (77 μm thickness) with only 2.8 vol. % of quasi-1D $TaSe_3$ exfoliated atomic thread fillers delivered 15 dB of total EMI shielding in the practically important X-band GHz frequency range. The EMI shielding efficiency of the developed materials expressed via the total shielding efficiency normalized by the mass density, thickness and filler loading fraction, $Z_B\sim220$ $dB/cm^2g$, exceeds that of other polymers with various metallic, carbon nanotube or graphene fillers. The EMI shielding performance of the films with the quasi-1D fillers in EHF-band of sub-THz frequencies was particularly impressive. Total shielding efficiency $SE_T$ changed from 60 dB to above 70 dB as the frequency varied from 240 GHz to 320 GHz. This performance was achieved in composite films with only 1.3 vol. % loading of exfoliated quasi-1D fillers of $TaSe_3$ and the film thickness of 1 mm.  Interestingly, the efficient EMI shielding was achieved in polymer films, which retained their DC electrically insulating properties, essential for many applications. The developed polymer films with quasi-1D fillers are promising for 5G-and-beyond communication technologies, which require electromagnetic shielding films, which are flexible, light-weight, corrosion resistive, electrically insulating and inexpensive.

## METHODS

**Preparation of TaSe$_3$:** 1.7315 g (9.57 mmol) of tantalum (STREM 99.98% purity) and 2.2718 g (28.8 mmol) of selenium (STREM 99.99% purity) were ground together gently with an agate mortar/pestle. This mixture was added to a 17.78 × 1 cm fused quartz ampule along with 62.3 mg iodine, (J.T. Baker, 99.9% purity). The ampule was evacuated and backfilled with Ar three times while submerged in an acetonitrile/dry ice bath, then flame sealed under vacuum. The ampule was placed in a Carbolite EZS 12/450B three-zone horizontal tube furnace and heated to 750 – 650 °C (source zone – growth zone) for 336 h. After the ampule had cooled to room temperature and was opened, the isolated shiny black crystals were left to sit in a fume hood for 1-2 h to allow excess iodine to sublime.

**Polymer Composite Preparation:** The bulk TaSe$_3$ crystals were added to acetone with a starting concentration of 0.5 mg/mL in 10 mL cylindrical vials and sonicated in a low power sonic bath (Branson 5510) for several hours. The vials were inspected visually every 2 h to verify the quality of the dispersion. The resultant dispersion was centrifuged (Eppendorf Centrifuge 5810) at 7000 rpm for 5 to 10 min. The supernatant was collected and poured in a Peltier dish to dry for characterization purposes. The precipitate, as well as some material stuck to the side walls of the vial after centrifugation, was collected and left in the ambient air until the solvent evaporated. The resulting dark brown, exfoliated TaSe$_3$ threads exhibited different aspect ratios (see Figure S3, S4 for more SEM images in Supplementary Information). The variation in aspect ratio of the TaSe$_3$ fillers is beneficial in EMI shielding applications and has been discussed in the text. The obtained fillers were mixed in precalculated proportions with three different off-the-shelf base polymeric matrices of UVP, SA, and epoxy. The UVP was mixed with low volume fraction of TaSe$_3$ at 500 rpm for 10 min using a high-shear speed mixer (Flacktek Inc.). The prepared mixture was sandwiched between two pieces of nylons and pressed gently until a thin film formed in between. The sandwich was left under the UV light for 2 min to cure. After that, the nylons were separated easily and a flexible film of UVP-TaSe$_3$ remained, as shown in Figure 1f. In case of SA-based flexible films, the SA powder was added to the DI water, sealed and stirred for 2 hours on top of a hot plate with temperature set to 50 °C Then, the TaSe$_3$ filler was added to the solution at low concentrations. The mixture was stirred and sonicated for 30 min and drop cast on a Peltier dish. The dish was placed on a hot plate at 50 °C for almost 1 h. The resultant was a dark brown flexible film shown in Figure 1g. The epoxy composites were made by mixing the epoxy resin (bisphenol-A-(epichlorhydrin), molecular weight ≤700, Allied High-Tech Products, Inc.) and hardener (triethylenetetramine, Allied High-Tech Products, Inc.) with the mass ratio of 100 to 12, respectively. The TaSe₃ filler was added afterwards and mixed with the high-shear speed mixer at 500 rpm for 10 min. The compound was vacuumed for 10 min to remove the possible trapped air bubbles. The compound was mixed one more time at 300 rpm for 10 min, vacuumed, and then poured into special molds to cure. The product is the dark composite shown in Figure 1h. More details of the sample preparation are provided in the Supplementary Information.

**Mass Density Measurements:** Using an electronic scale (Mettler Toledo), the weight of the samples was measured in air ($w_a$) and in DI water ($w_w$). In case of SA flexible films, the weights of the films were measured in air and ethanol ($w_e$) since SA is soluble in DI water. The mass density of the samples were calculated according to Archimedes principle $\rho_c = (w_a/(w_a - w_{w,e})) \times (\rho_{w,e} - \rho_a) + \rho_a$ where $\rho$ is the density and subscripts "$a$", "$w$", and "$e$" corresponds to air, water, and ethanol, respectively. The volume fraction, $\phi$, of the TaSe₃ filler was calculated according to the rule of mixtures as $\phi = (\rho_c - \rho_p)/(\rho_f - \rho_p)$ where $\rho_p$ and $\rho_f$ are the density of the base polymer and TaSe₃ filler, respectively. The density values of each sample with its constituents are listed in Supplementary Information.

**Electromagnetic Interference Shielding Measurements in X-Band:** To determine EMI characteristics, we measured the scattering parameters, $S_{ij}$, using the two-port PNA system. The indices $i$ and $j$ represents the ports, which are receiving and emitting the EM waves. Each port can simultaneously emit and detect the EM waves and thus, the results of the measurements include four parameters of $S_{11}$, $S_{12}$, $S_{21}$, and $S_{22}$. Owing to the symmetry of the samples, one can expect that $|S_{11}| = |S_{22}|$ and $|S_{12}| = |S_{21}|$. The scattering parameters are related to the coefficients of reflection, $R = |S_{11}|^2$, and transmission, $T = |S_{21}|^2$. The measurements were performed in the X-Band frequency range (8.2 – 12.4 GHz) with the frequency resolution of 3 MHz. A Programmable Network Analyzer (PNA) Keysight N5221A was used. The PNA was calibrated for 2-port measurements of scattering parameters $S_{ij}$ at input power $P_{in} = 3$ dBm. A WR-90 commercial grade straight waveguide with two adapters at both ends with SMA coaxial ports was used as a sample holder. Special cables were used for high temperature RF measurements. The samples were a bit larger than the rectangular cross section (22.8×10.1 mm²) of the central hollow part of the waveguide in order to prevent the leakage of EM waves from the sender to receiver antenna. The scattering parameters, $S_{ij}$, were directly measured and used to extract the reflection and absorption shielding efficiency of the composites.

**Electromagnetic Interference Shielding Measurements in EHF-Band:** Due to a small cross section of the WR-3 waveguide, EMI characteristics in the sub-THz range were measured in free space. One of the most commonly used free-space techniques at THz and sub-THz frequencies is the time-domain spectroscopy (THz-TDS).[57] Its efficiency is limited to frequencies below 300 GHz due to the low power of the excitation signal in this spectral range.[58] Characterization of highly absorptive materials using THz-TDS may not be feasible. For this reason, the EMI shielding efficiency was determined from the measured scattering parameters using Agilent N5245A vector network analyzer (VNA) with a pair of frequency extenders from Virginia Diodes Inc.[31,59] The quasi-optical path of the measurement setup consisted of two high-gain horn antennas and two double convex lenses to focus the EM wave on the sample under test. The measurements were performed in the frequency range from 220 GHz to 320 GHz. The VNA with frequency extenders was calibrated using the Thru – Reflect – Line (TRL) method. The reference planes for 2-port measurements were achieved at the ends of the waveguide ports of both extenders. To compensate for the transmission losses in the measurement path two additional reference measurements were performed. The measurement with an empty optical path and the measurement with a metal plate allow one to calculate the actual transmission and reflection coefficients, respectively. In order to improve the reliability of the collected data additional time-domain gating was applied (see [31,59–61] for more details).

## Acknowledgements

The work in Balandin group was supported, in part, by the National Science Foundation (NSF) program Designing Materials to Revolutionize and Engineer our Future (DMREF) via a project DMR-1921958 entitled Collaborative Research: Data Driven Discovery of Synthesis Pathways and Distinguishing Electronic Phenomena of 1D van der Waals Bonded Solids; the Semiconductor Research Corporation (SRC) contract 2018-NM-2796 entitled One-Dimensional Single-Crystal van-der-Waals Metals: Ultimately-Downscaled Interconnects with Exceptional Current-Carrying Capacity and Reliability; and by the Proof of Concept (POC) project of the Office of Technology

Partnerships (OTP), University of California, Riverside (UCR). S.R. and Y.Y. acknowledge support from the CENTERA Laboratories in the framework of the International Research Agendas Program for the Foundation for Polish Sciences co-financed by the European Union under the European Regional Development Fund (no. MAB/2018/9). Z.B. and F.K. thank Amirmahdi Mohammadzadeh and Dylan Wright, POEM Center, UCR for assistance with SEM characterization.

## Contributions

A.A.B. and F.K. conceived the idea of the electromagnetic shielding films with quasi-1D fillers and planned the study. A.A.B. and F.K. coordinated the project and contributed to the experimental data analysis. Z.B. exfoliated the material, prepared the composites, performed measurements in the X-band frequency range, and analyzed the experimental data. Y.G. synthesized bulk crystals and conducted material characterization. S.G. conducted Raman spectroscopy of exfoliated materials and thin films. S.S. conducted SEM characterization. T.T.S. supervised material synthesis and contributed to data analysis. A.A.B. and F.K. led the manuscript preparation. Y.Y. and K.G. carried out measurements in EHF frequency range and processed experimental data using the time-domain gaiting method. G.C. and S.R. contributed to data analysis and interpretation. All authors contributed to writing and editing of the manuscript.

*Commun. MIKON 2016*, Institute Of Electrical And Electronics Engineers Inc., **2016**.

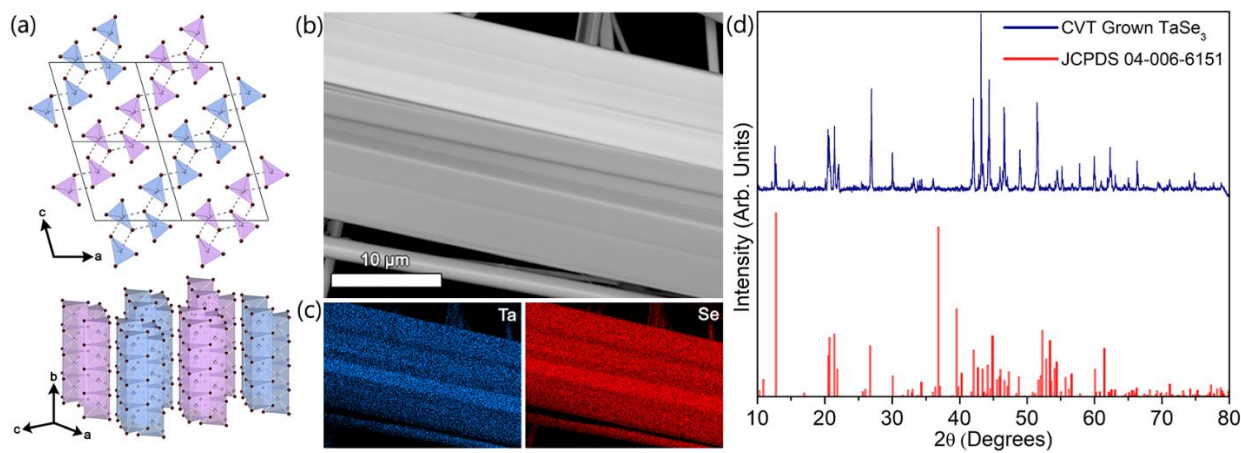

**Figure 1: Structure and composition of as-synthesized TaSe₃.** (a) Crystal structure of TaSe₃; red spheres represent Se and blue/purple spheres represent Ta. The parallelograms in the top panel outline unit cells viewed along the *b*-axis, perpendicular to the TaSe₃ chains. The side view in the bottom panel shows the quasi-1D nature of trigonal prismatic [TaSe₆] units extending along the *b*-axis. The corrugated bilayer nature of this structure is emphasized with the Ta···Se interchain interactions and the purple/blue coloring; bilayers are separated from their neighbors by van der Waals gaps. (b) SEM image of a mechanically-exfoliated TaSe₃ crystal. (c) Corresponding EDS mapping showing uniform overlap of Ta and Se along the length of the crystal. (d) Powder XRD pattern of CVT-grown TaSe₃ crystals (blue) matching a reference pattern of monoclinic TaSe₃ (red).

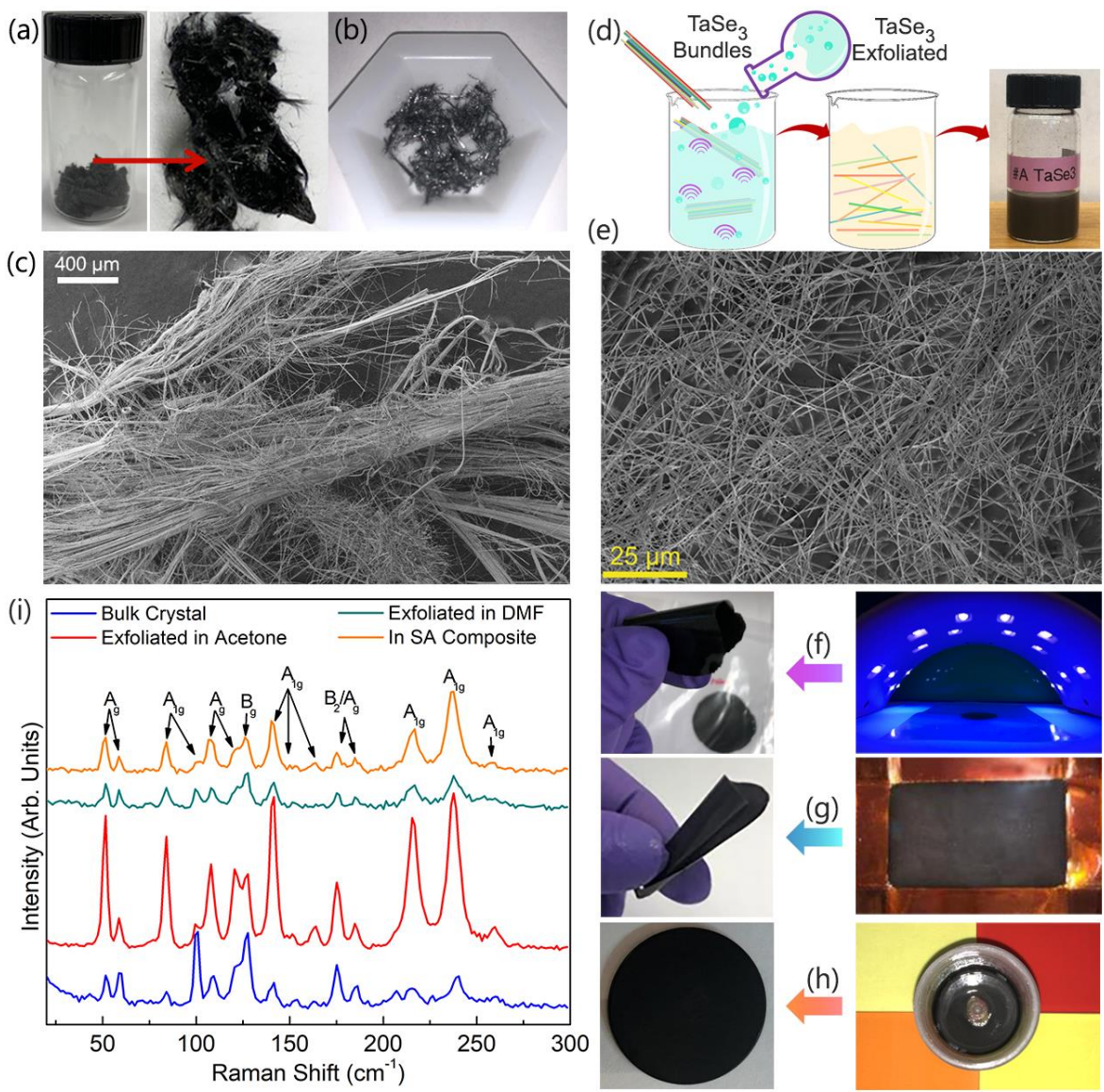

**Figure 2: Composite films preparation and characterization**. (a, b) As-prepared TaSe₃ crystals and fibers. (c) SEM image of fibrous TaSe₃ bundles. (d) Schematic showing the process of chemical liquid-phase (LPE) exfoliation using low-power bath sonication. The vial contains exfoliated TaSe₃ in acetone. (e) SEM image of TaSe₃ threads after liquid phase exfoliation in acetone. Note the high aspect ratio morphologies. (f, g) Synthesis of flexible polymeric films using a special UV-cured polymer (f) and sodium alginate (g) as the matrix and exfoliated TaSe₃ as filler. (h) Optical image of the epoxy composite containing about 1.3 vol. % concentration of exfoliated TaSe₃ threads. (i) Raman spectra of the TaSe₃ before (blue) and after solvent-assisted exfoliation in acetone (red) and DMF (cyan). The orange curve shows the Raman spectrum of the TaSe₃ filler mixed with sodium alginate (SA). The characteristic Raman peaks of TaSe₃ do not change as it is exfoliated or combined with SA polymer.

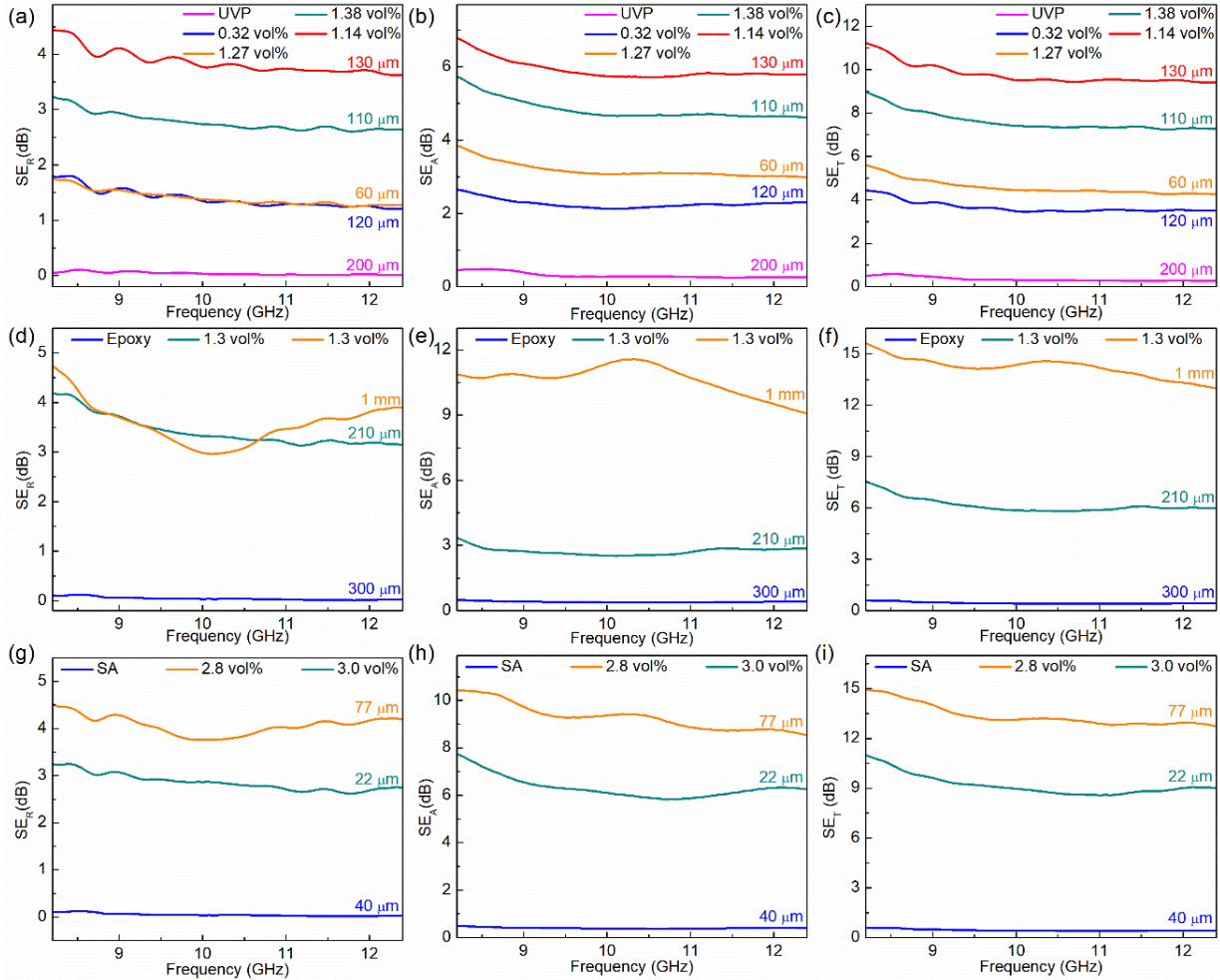

**Figure 3: Electromagnetic characteristics of films with low concentration of quasi-1D TaSe₃ fillers in X-band frequency range.** Reflection (SE$_R$), absorption (SE$_A$), and total (SE$_T$) electromagnetic interference shielding efficiency of (a-c) UV-cured polymer (d-f) epoxy, and (g-i) sodium alginate films and composites with low concentration of quasi-1D TaSe₃ bundles of atomic threads as fillers. The concentration is indicated in the legends. Note that only 1.3 vol. % of quasi-1D fillers can provide ~15 dB shielding efficiency, SE$_T$, in the electrically insulating films (for reference, SE$_T$=10 dB corresponds to blocking 90% of electromagnetic energy).

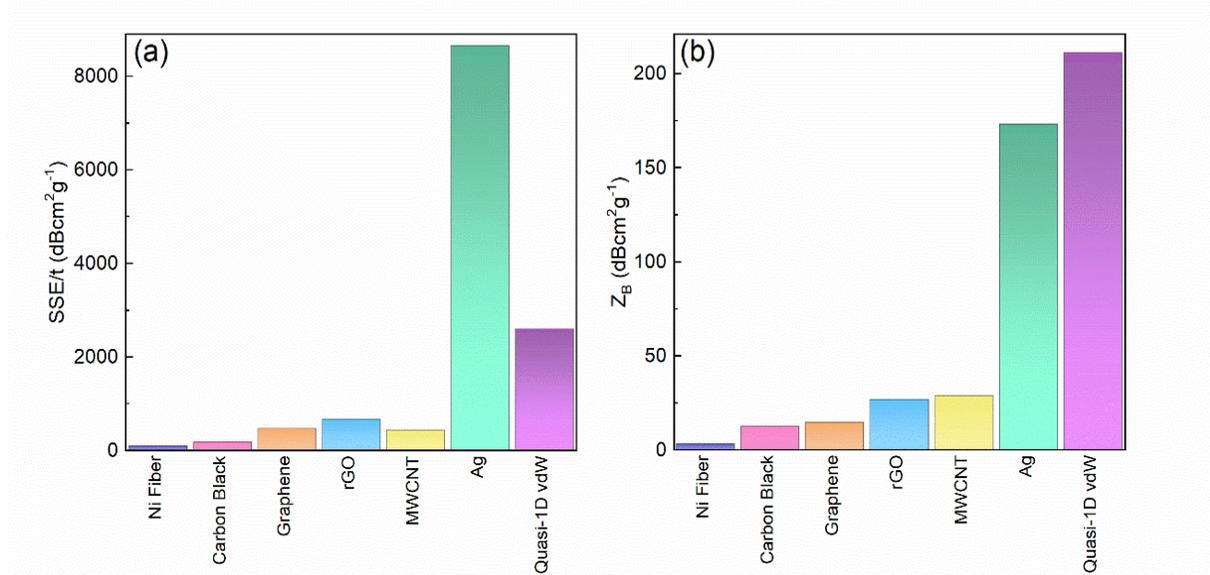

**Figure 4: EMI shielding characteristics of polymeric composites with different fillers.** (a) The specific EMI shielding efficiency normalized by thickness. A polymer composite with 90 wt. % of Ag inclusion exhibit the highest $SSE/t$. (b) The same plot in panel (a) normalized by the filler weight loading fraction. The $Z_B$ factor indicates composite's shielding effectiveness per aerial density of the filler. The lower the thickness, density, and filler weight loading fraction and higher the total shielding effectiveness, the higher the $Z_B$.

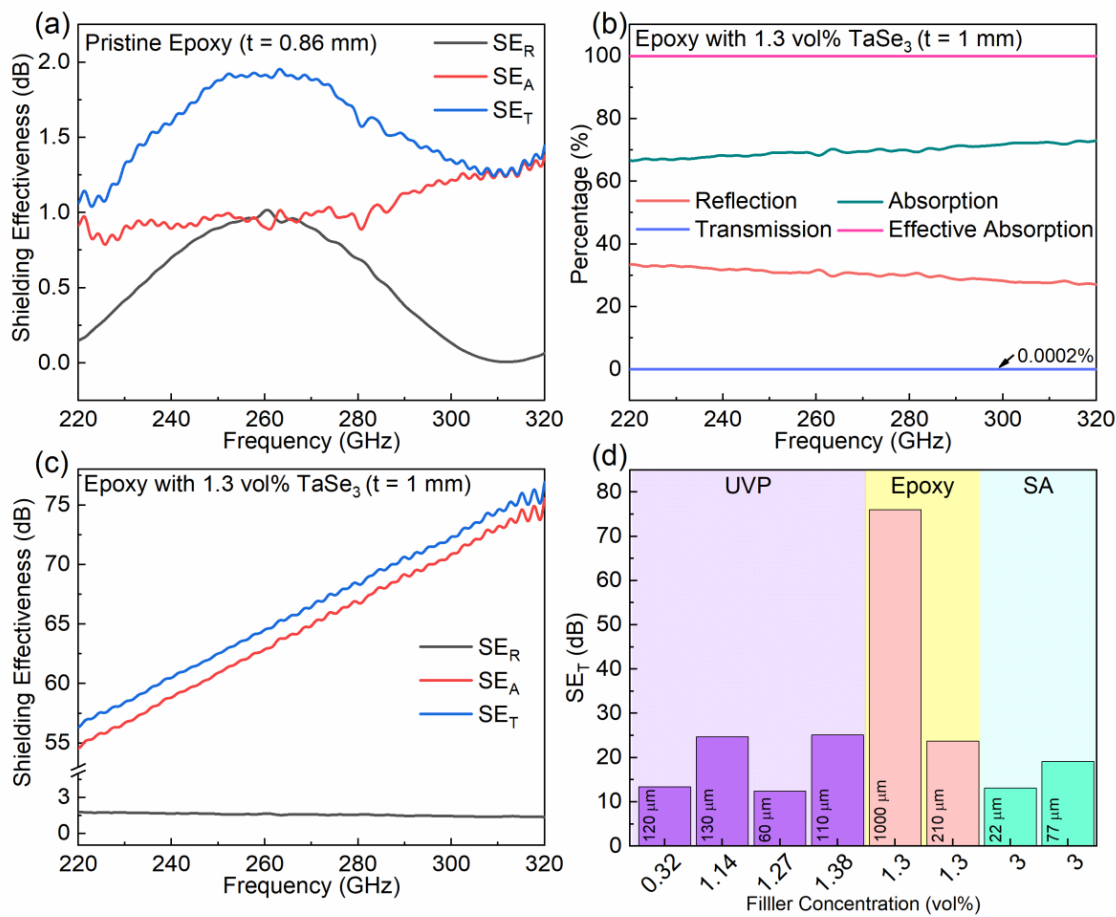

**Figure 5: Electromagnetic shielding characteristics of the films with low concentration of quasi-1D TaSe₃ fillers in EHF band.** (a) Shielding effectiveness of pristine epoxy used as the base material for some of the composites. (b) Reflection, absorption, effective absorption, and transmission coefficients of epoxy with 1.3 vol. % loading of the quasi-1D TaSe₃ fillers. Note that in the EHF range, almost all the incident EM wave energy is blocked and only 0.0002% is transmitted. (C) Reflection, absorption, and total shielding effectiveness of the same composite. Note that absorption is the dominant mechanism in blocking the EM waves in EHF band. (d) Total shielding effectiveness of all samples tested in the EHF band. The results are shown for the frequency 320 GHz. The total shielding effectiveness scales with the loading fraction of quasi-1D fillers and the thickness of the films.